\documentclass[fleqn,usenatbib]{mnras}

\usepackage{newtxtext,newtxmath}
\usepackage{bm}
\usepackage{threeparttable}
\usepackage{graphicx,times}   
\usepackage{natbib}
\usepackage{amsmath}
\usepackage[T1]{fontenc}
\usepackage{siunitx}

\DeclareRobustCommand{\VAN}[3]{#2}
\let\VANthebibliography\thebibliography
\def\thebibliography{\DeclareRobustCommand{\VAN}[3]{##3}\VANthebibliography}

\usepackage{graphicx}	% Including figure files
\usepackage{amsmath}	% Advanced maths commands
\title[Radial grain size distribution in the DS Tau disk]{The Radial Profile of Dust Grain Size in the Protoplanetary Disk of DS\,Tau}

\author[Dafa Li et al.]{
Dafa Li,$^{1,2}$
Yao Liu,$^{1}$\thanks{email:yliu@pmo.ac.cn}
Hongchi Wang,$^{1, 2}$
Yao Wang$^{1}$
and Yuehui Ma$^{1}$
\\
$^{1}$Purple Mountain Observatory \& Key Laboratory of Radio Astronomy, Chinese Academy of Sciences,
10 Yuanhua Road, Qixia District, Nanjing 210033, P. R. China;\\
$^{2}$School of Astronomy and Space Science, University of Science and Technology of China, 96 Jinzhai
Road, Hefei 230026, P. R. China\\
}

\date{Accepted XXX. Received YYY; in original form ZZZ}

\pubyear{2022}

\begin{document}
\label{firstpage}
\pagerange{\pageref{firstpage}--\pageref{lastpage}}
\maketitle

\begin{abstract}
How do dust grains in protoplanetary disks overcome rapid radial drift and grow from micron size particles to planets is not well understood. The key is to search for evidence of dust accumulation and growth as a function of radius in the disk. We investigate the radial profile of grain size in the DS\,Tau disk by fitting multi-band ALMA observations with self-consistent radiative transfer models. The best-fit grain sizes range from centimeters in the inner disk down to ${\sim}\,30\,\mu{\rm m}$ in the outer regions. Such an inside-out decreasing tendency is consistent with theories of dust evolution. Based on the best-fit model, we find that dust of ${\sim}\,2$ Jupiter masses has been depleted within the gap. By taking the gas-to-dust mass ratio into account, the lost mass is enough to form the 3.5 Jupiter mass planet inferred by literature hydrodynamic simulations.  Moreover, our modeling also indicates that at the interface region between the gap and the ring, the grain size profile shows a discontinuity, with its amplitude dependent on the dust model adopted in the radiative transfer analysis. Future multi-wavelength observations at higher angular resolutions are required to better constrain the grain size and its variation in the vicinity of disk substructures.
\end{abstract}


%%%%%%%%%%%%%%%%%%%%%%%%%%%%%%%%%%%%%%%%%%%%%%%%%%

%%%%%%%%%%%%%%%%% BODY OF PAPER %%%%%%%%%%%%%%%%%%

\section{Introduction}
\label{sect:intro}
Protoplanetary disks around young stars, composed of dense gas and dust, provide raw materials for the formation of planets \citep{2011ARA&A..49...67W}. During the process of planet formation, the size of dust particles will increase by more than 13 orders of magnitude. Dust growth starts from the sticking and coagulation of micron-sized grains to the formation of millimeter pebbles, see \citet{2014prpl.conf..339T} for a review.  Observations at (sub-)millimeter wavelengths are an essential tool to investigate the first step of planet formation, because such data trace the distribution of millimeter grains in the disk midplane.\par

Dust continuum emission in the (sub-)millimeter wavelength regime roughly follow a power law $F_{\nu}\,{\propto}\,\nu^{\alpha_{\mathrm {mm}}}$, where $\nu$ is the emission 
frequency, and $F_{\nu}$ is the flux density. When multi-band data are available, the spectral index $\alpha_{\mathrm {mm}}$ can be measured, and it is frequently used as a probe of 
grain size \citep{1991ApJ...381..250B,2005ApJ...631.1134A,2007ApJ...671.1800A,2010A&A...512A..15R,2010A&A...521A..66R,2012ApJ...760L..17P,2016A&A...588A..53T,2018ApJ...859...21A}. At (sub-)millimeter wavelengths, the mass absorption 
coefficient ($\kappa_{\nu}$) of dust grains can be approximated as $\kappa_{\nu}\,{\propto}\,\nu^{\beta_{\mathrm {mm}}}$. For micron-sized solids, 
the slope $\beta_{\mathrm {mm}}$ does not change with dust size at first (for grain size $\lesssim\,20\,\mu{\rm m}$), and then increases with the size (for  $20\,\mu{\rm m}\,{\lesssim}$ grain size ${\lesssim}\,300\,\mu{\rm m}$). When reaching a specific size (for grain size $\gtrsim\,300\,\mu{\rm m}$), $\beta_{\mathrm {mm}}$ decreases with size \citep{2001ApJ...553..321D,2004A&A...416..179N,2006ApJ...636.1114D}. The $\beta_{\mathrm {mm}}$ value for millimeter pebbles is obviously lower than that of micron-sized particles (${\sim}\,1.7$, \citet{Li2001}), see Figure 8 in \citet{2020ARA&A..58..483A}. In the optically thin cases, the measured spectral index $\alpha_{\mathrm {mm}}$ is linked to $\beta_{\mathrm {mm}}$ via $\beta_{\mathrm {mm}}\,{=}\,\alpha_{\mathrm {mm}}\,{-}\,\log (B_{\nu 1} / B_{\nu 2}) / \log (\nu_{1} / \nu_{2})$, where $B_{\nu}$ is the Planck function. Therefore, the millimeter spectral index tracks the grain size by its relationship with the slope of dust absorption coefficient.\par

When the disk is spatially resolved at multi-wavelengths, one can obtain the variation in $\alpha_{\mathrm {mm}}$ along the radial direction, enabling the investigation on the grain size as a function of radius ($R$). Previous works based on the Submillimeter Array, Combined Array for Research in Millimeter-wave Astronomy and Very Large Array data found that the maximum grain size, $a_{\rm max}$, generally follows a power-law form, $a_{\max}(R)\,{\propto}\,R^{-b}$, with $b\,{\sim}\,1-2$ \citep[e.g.,][]{Trotta2013,2016A&A...588A..53T}. Nevertheless, limited by the spatial resolutions of these early generation interferometers, typically ${\sim}\,0.3^{\prime\prime}$, properties of dust grains in the inner disk regions are difficult to be constrained. The situation has been significantly improved by the advent of the Atacama Large Millimeter/submillimeter Array (ALMA),  with spatial resolution being even better than $0.05^{\prime\prime}$. Recent ALMA observations have revealed the previously undiscovered substructures in protoplanetary disks, including rings-cavities \citep{2007A&A...467..163P, 2009ApJ...704..496B,2011ApJ...732...42A}, rings-gaps \citep{2018ApJ...859...21A,2018ApJ...869...17L}, spirals \citep{2016Sci...353.1519P,2018ApJ...869L..43H}, arcs \citep{2013Natur.493..191C,2013ApJ...775...30I,2015ApJ...798L..44M,2018ApJ...869L..49I}, and vortices \citep{2009ApJ...700.1502A,2020A&A...641A.128R}. The formation mechanisms of these substructures are still in debate. Proposed scenarios include fluid instabilities, dynamical interactions between disk and planetary companions, and condensation fronts, see \citet{2020ARA&A..58..483A} for a review. Analyzing the dust properties, for instance the composition and especially grain size, and their correlation with the substructures, may help to identify the mechanisms at work. 

\begin{table}
 \centering
 \caption{Flux measurements of DS\,Tau}  
 \doublerulesep 0.1pt \tabcolsep 3pt %space between two columns. 
     \begin{tabular}[h]{rrrrr}
     \hline
Wavelength & Photometry & Error of the photometry & Filter & References \\
 $[\mu{\rm m}]$      &  [mJy]  &  [mJy]   &    &  \\
\hline\noalign{\smallskip}
    0.36  & $10.81$ & $4.88$& U &1 \\
    0.44  & $19.50$ & $7.72$ & B & 1 \\
    0.51  & $24.45$ & $1.52$ & $\mathrm G_{\mathrm {BP}}$ & 2 \\
    0.55  & $43.24$ & $16.32$ & V & 1 \\
    0.62  & $46.28$ & $0.76$ & G & 2 \\
    0.64  & $73.21$ & $28.31$ & $\mathrm R_{\mathrm c}$ & 1 \\
    0.77  & $91.31$ & $3.75$ & $\mathrm G_{\mathrm {RP}}$ & 2 \\
    0.79  & $122.05$ & $48.32$ & $\mathrm I_{\mathrm c}$ & 1 \\
    1.24  & $260.91$ & $4.32$ & J & 3 \\
    1.66  & $372.82$ & $11.33$ & H & 3 \\
    2.16  & $406.94$ & $10.87$ & $\mathrm K_{\mathrm s}$ & 3 \\
    3.36  & $355.40$ & $9.49$ & WISE & 4 \\
    3.55  & $315.69$ & $5.81$ & IRAC & 5 \\
    4.49  & $309.08$ & $5.69$ & IRAC & 5 \\
    4.60  & $315.49$ & $6.10$ & WISE & 4 \\
    5.73  & $255.09$ & $7.05$ & IRAC & 5 \\
    7.87  & $272.92$ & $7.54$ & IRAC & 5 \\
    11.56  & $262.48$ & $3.38$ & WISE & 4 \\
    22.08  & $292.12$ & $6.72$ & WISE & 4 \\
    70.00  & $200.00$ & $40.00$ & PACS & 6 \\
    100.00  & $280.00$ & $60.00$ & PACS & 6 \\
    160.00  & $240.00$ & $50.00$ & PACS & 6 \\
    869.00  & $39.00$ & $4.00$ & SCUBA & 7 \\
    1250.00  & $25.00$ & $6.00$& IRAM & 8 \\
    1330.00  & $22.24 $ & $0.10$ & ALMA & 9 \\
    2900.00  & $2.90$ & $0.03$ & ALMA & 10 \\
    2974.00  & $2.96$ & $0.21$ & PdBI & 11 \\
  \noalign{\smallskip}\hline
    \end{tabular}
    \begin{tablenotes}
\footnotesize
\item
(1) \cite{1995ApJS..101..117K}, (2) \cite{2018yCat.1345....0G}, (3) \cite{2003yCat.2246....0C}, (4) \cite{2012yCat.2311....0C}, (5) \cite{2010ApJS..186..111L}, (6) \cite{2017ApJ...849...63R}, (7) \cite{2005ApJ...631.1134A}, (8) \cite{1995ApJ...439..288O}, (9) \cite{2018ApJ...869...17L}, (10) \cite{2020ApJ...898...36L}, (11) \cite{2010A&A...512A..15R}.
\end{tablenotes}
    \label{tab:sed}
\end{table}

\citet{2017A&A...607A..74L} analyzed the ALMA and VLA observations of the HL\,Tau disk with self-consistent radiative transfer models. By fixing the maximum grain size $a_{\rm max}$ to 3\,mm, they explored the grain size distribution ${N(a)}\,{\propto}\,a^{-p}da$, and found that the slope $p$ in the bright rings is generally smaller than that in the dark gaps, which implies that large dust grains are more concentrated in the rings. Instead, \citet{2021ApJS..257...14S} fixed the grain size slope $p$ to 2.5, and fitted the ALMA band 6 ($\lambda\,{\sim}\,1.3\,\rm{mm}$) and band 3 ($\lambda\,{\sim}\,2.9\,\rm{mm}$) data for five other disks by adjusting the maximum grain size $a_{\rm max}$ at each radius. Their results show that $a_{\rm max}$ locally peaks in most of the known rings. It should be noted that their models are based on approximate solutions to the radiative transfer equation under certain assumptions. For instance, the dust temperature and grain size in the radial direction are independent on each other, and the vertical structure of the disk is not taken into account. How, and to which degree these simplifications affect the results need to be investigated.\par

Our research target, DS\,Tau, is an M0.4 young star \citep{2014ApJ...786...97H} in the Taurus star-forming region at a distance of 158\,pc \citep{2021A&A...649A...1G}. It was observed with ALMA at Band 6 and 3 through the Cycle 4 (ID: 2016.1.01164.S; PI: Gregory Herczeg) and Cycle 6 program (ID: 2018.1.00614.S; PI: Feng Long), respectively. Thanks to the high-resolution observations (\textasciitilde\,0.1\,arcsec, corresponding to\,\textasciitilde\,16\,AU), substructures, including a gap (width\,\textasciitilde\,27\,AU, centred at\,\textasciitilde\,33\,AU) and a ring (width\,\textasciitilde\,17\,AU, centred at\,\textasciitilde\,56\,AU), are revealed \citep{2018ApJ...869...17L,2020ApJ...898...36L}. \citet{2019MNRAS.486..453L} analyzed the width of the gap and suggested a 5.6 Jupiter mass planet embedded in the gap is responsible for the ring formation. Based on hydrodynamic simulations, \citet{2020MNRAS.495.1913V} suggested the mass of the planet is $M_{\mathrm P}\,{=}\,3.5\,{\pm}\,1\,\mathrm{M_{\mathrm {Jup}}}$. However, this conclusion needs to be verified with future observations. In this paper, we build a self-consistent radiative transfer model to fit the spectral energy distribution (SED) and ALMA images at 1.3 and 2.9\,mm to explore the growth and distribution of dust grains, and their relationship with the substructures. The observational data are presented in Section~\ref{sect:obs}. Section~\ref{sect:model} and \ref{sect:Fit} describe the model setup and fitting approach, respectively. In Section~\ref{sect:discussion}, we present and discuss our results. A summary is given in Section~\ref{sect:summary}.\par

\begin{figure}
\centering
\includegraphics[width=\columnwidth, angle=0]{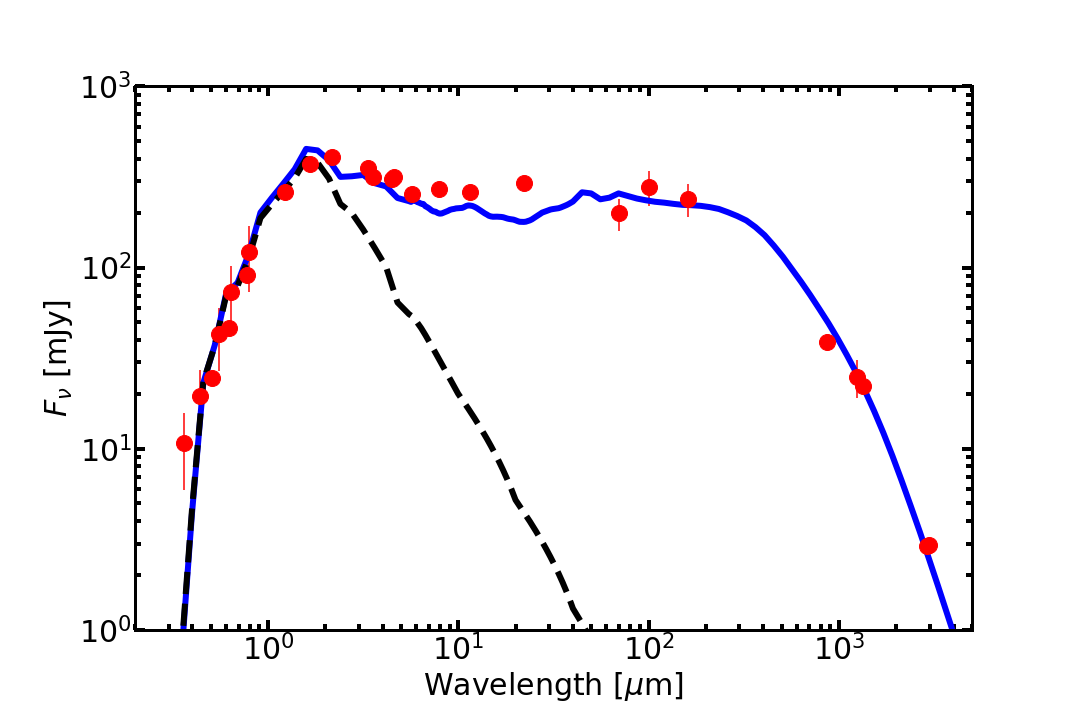}
\caption{Spectral energy distribution (SED) of DS\,Tau. The red dots with error bars represent the observational data (see Table~\ref{tab:sed}). The blue line shows our best-fit model. Black dashed line is the input stellar spectrum.}
\label{sed}
\end{figure}

\section{Observations}
\label{sect:obs}
Table~\ref{tab:sed} lists the measured flux densities of DS\,Tau, and the data points cover a broad range of wavelength from the optical to millimeter. The observations of \citet{2018yCat.1345....0G} and \citet{1995ApJS..101..117K} provide the optical photometry. The infrared data provided by \citet{2003yCat.2246....0C}, \citet{2012yCat.2311....0C}, \citet{2010ApJS..186..111L} and \citet{2017ApJ...849...63R} are essential to determine the evolutionary stage of the disk. (Sub-)millimeter data, taken from  \citet{2005ApJ...631.1134A}, \citet{1995ApJ...439..288O}, \citet{2018ApJ...869...17L}, \citet{2020ApJ...898...36L} and \citet{2010A&A...512A..15R}, can constrain the dust mass 
and grain size distribution in the disk. Figure~\ref{sed} shows the SED (red dots) along with our best-fit model SED (blue line). \par
 
\begin{figure}
    \centering
    \includegraphics[width=0.9\columnwidth,angle=0]{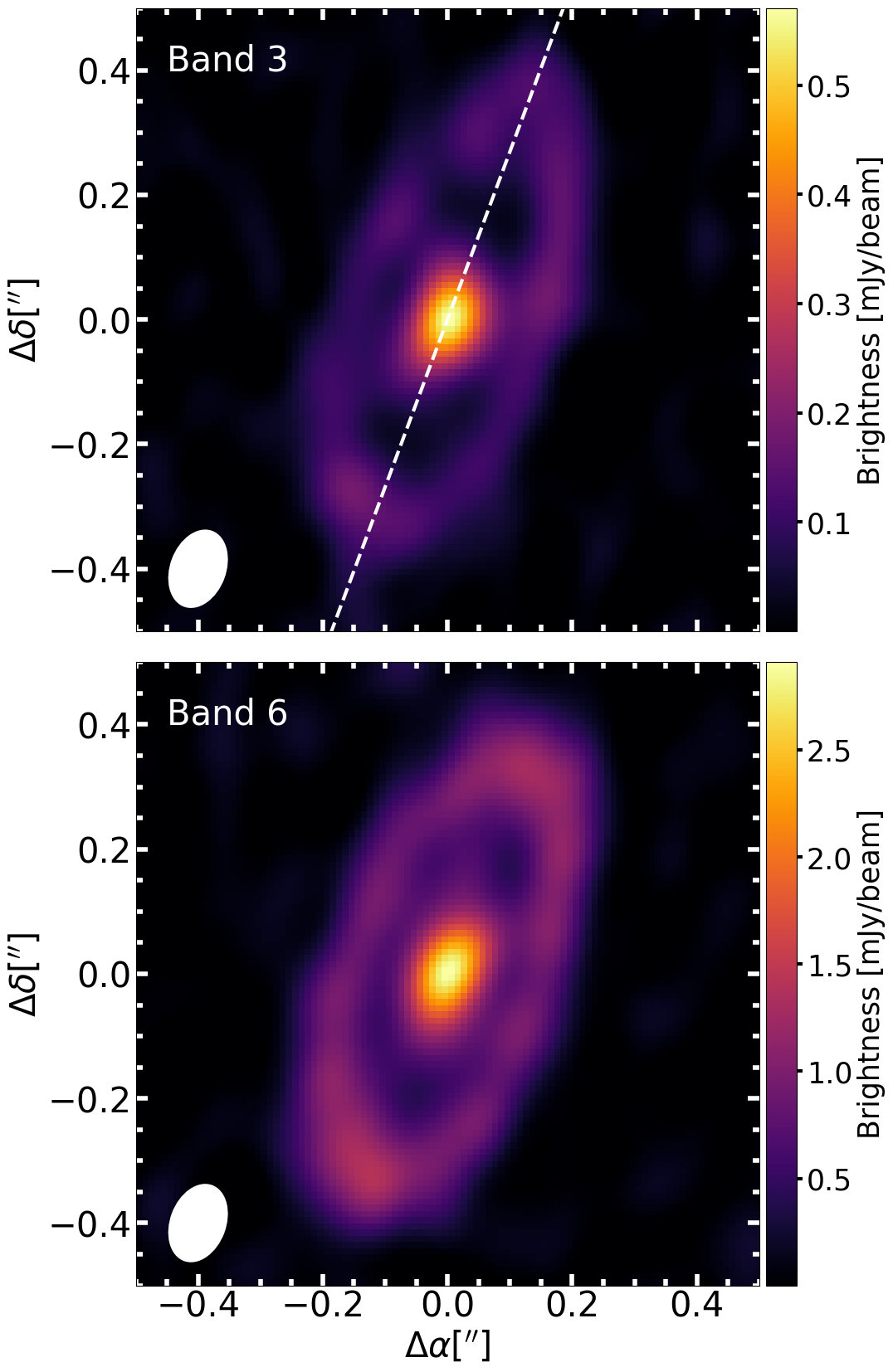}
\caption{ALMA continuum images at band 3 (2.9\,mm; top panel) and band 6 (1.3\,mm; bottom panel). The white dashed line represents the major axis of the disk, with a $\mathrm {PA}\,{=}\,159.62^{\circ}$ \citep{2018ApJ...869...17L,2020ApJ...898...36L}.}
\label{fig:almaimg}
\end{figure}

\begin{figure*}
\centering
\includegraphics[width=0.8\textwidth,angle=0]{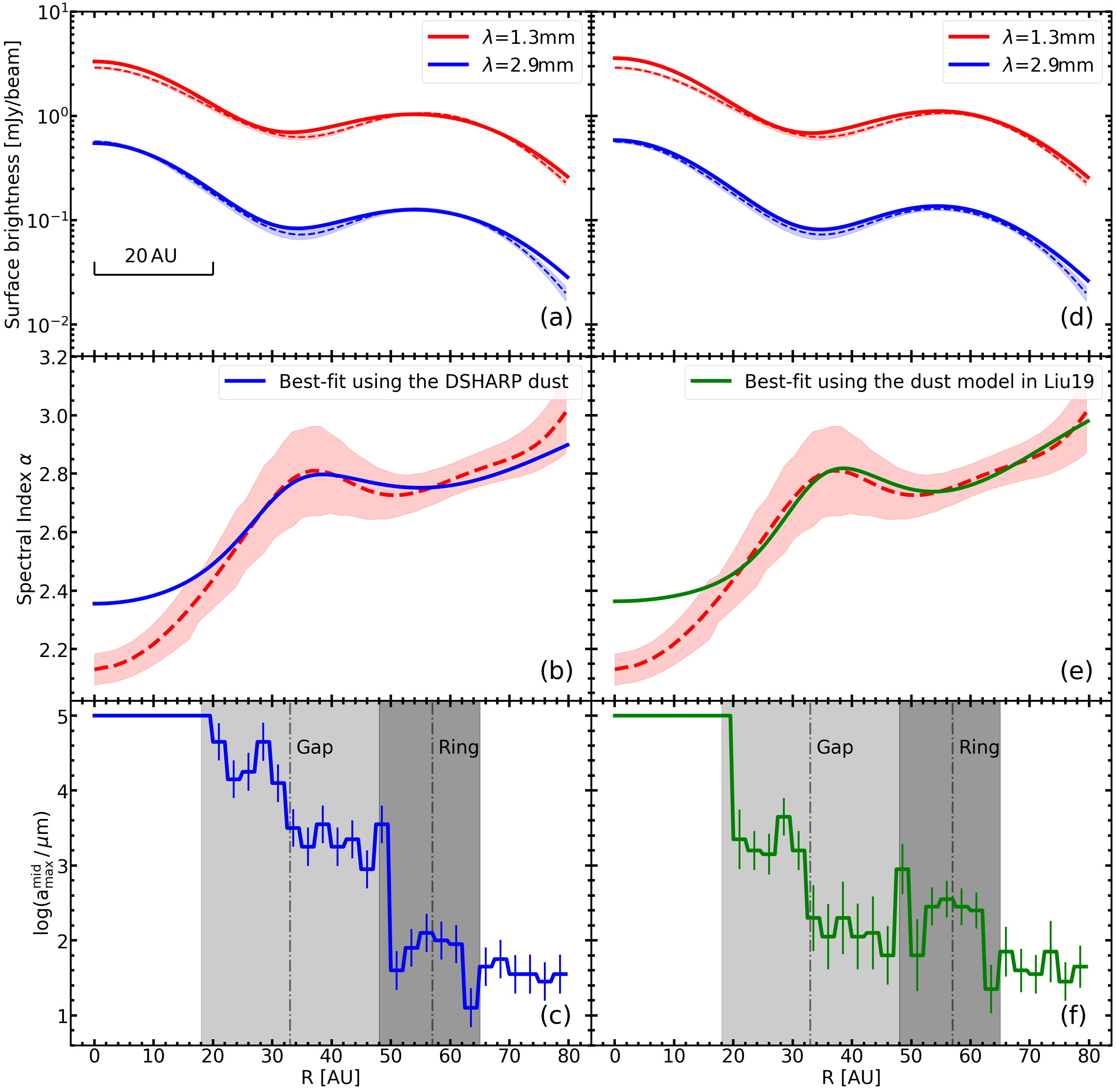}
\caption{Results of fitting the ALMA images. {\it Panel (a):} Radial brightness profiles at $\lambda=1.3\,\rm{mm}$ (red line) and $\lambda=2.9\,\rm{mm}$ (blue line). The observed profiles are shown with dashed lines, and their uncertainties are indicated with the shade areas. The solid lines refer to the best fit using the DSHARP dust model prescribed by \citet{2018ApJ...869L..45B}. {\it Panel (b):} Observed spectral index (red dashed line) and model prediction (blue solid line) using the DSHARP dust model. The red shaded area represents the uncertainty of the observation. {\it Panel (c):} Radial profile of the maximum grain size in the midplane ($a^{\mathrm {mid}}_{\mathrm {max}}$) of the best-fit model using the DSHARP dust opacities. The vertical blue solid lines represent the error of $a^{\mathrm {mid}}_{\mathrm {max}}$ in each radial bin. The two vertical dashed-dotted lines from left to right denote the positions of the gap and ring, with the gray shaded regions indicating their widths. {\it Panels (d)$-$(f):} Same as in {\it panels (a)$-$(c)}, but for the fitting results adopting a different dust model that is composed of 75\% amorphous silicate and 25\% carbon \citep{Liuy2019}.}
\label{fig:bestfit}
\end{figure*}

High resolution millimeter images are crucial for constraining the grain size distribution. Figure~\ref{fig:almaimg} shows the ALMA images at Band  3 and Band 6, both of which are convolved with 
the same beam of [$0.13^{\prime \prime},\,0.09^{\prime\prime}$] in size for the measurement of the spectral index. Details of the ALMA observations and data reduction can be found in \citet{2018ApJ...869...17L,2020ApJ...898...36L}. The millimeter spectral index as a function of $R$ is obtained from
\begin{equation}
\label{alpha}
\alpha_{\mathrm{mm}}(R)=\frac{\log [I_{\nu_{1}}(R) / I_{\nu_{2}}(R)]}{\log (\nu_{1} / \nu_{2})}
\end{equation}
where $I_{\nu_{1}}$ and $I_{\nu_{2}}$ are the radial intensity profiles at the two observed frequencies $\nu_{1}$ and $\nu_{2}$, respectively. The radial intensity profiles and the measured spectral index are shown in panels (a) and (b) of Figure~\ref{fig:bestfit}. We extract the intensity profiles by applying an azimuthal average that is performed on a series of concentric ellipses with shapes determined by the disk inclination and position angle, see Table~\ref{parameter}. The uncertainties are given by $\Delta{I_{\nu}}={\rm rms_{\nu}}/\sqrt{n}$, where $\rm{rms_{\nu}}$ is the rms noise in each map, and $n\,{=}\,L_{\rm ring}/B_{\rm maj}$ is the number of beams within an ellipse. The quantity $L_{\rm ring}$ is the circumference of the ellipse, and $B_{\rm maj}$ is the beam size in the major axis.

The absolute flux calibration accuracy has an impact on the uncertainty in the spectral index, especially when the spectral index is measured within one band \citep{Francis2020}. In our modeling, we do not consider this impact. Including the absolute flux calibration uncertainty, which is ${\sim}\,10\%$ at both ALMA bands \citep{2020ApJ...898...36L}, would introduce a systematic uncertainty of ${\sim}\,0.19$ in the spectral index. We show how the fitting results are influenced by this factor in Figure~\ref{fig:calibration}.

\section{Modeling setup}
\label{sect:model}
We develop a fitting method to constrain the radial variation of the grain size in the DS\,Tau disk with radiative transfer models that are simulated using the \texttt{RADMC-3D} \footnote{ \url{{https://www.ita.uni-heidelberg.de/~dullemond/software/radmc-3d/}}} code \citep{2012ascl.soft02015D}. Given the stellar parameters, dust density distribution and dust opacities, the code self-consistently computes the dust temperature using the Monte Carlo approach, and then simulates the SED and millimeter images. Previous works have shown that including dust scattering is important both in the estimation of disk dust mass and in the interpretation of observed spectral indices \citep[e.g.,][]{Zhu2019, Liuh2019,2021ApJS..257...14S}. Therefore, the option of isotropic scattering available in \texttt{RADMC-3D} is selected. In this section, we introduce the model setup including the density distribution, dust model and stellar properties.

\begin{table}
\centering
\caption[]{Parameter space of the model}\label{parameter}
\begin{tabular*}{\columnwidth}{@{}l@{\hspace*{10pt}}l@{\hspace*{10pt}}c@{\hspace*{10pt}}c@{\hspace*{10pt}}c@{\hspace*{10pt}}c}
\hline\noalign{\smallskip}
Parameter & Best fit & Min & Max & Grid points & Notes \\
\noalign{\smallskip}\hline
\multicolumn{6}{c}{Stellar parameters} \\
\noalign{\smallskip}\hline
$M_{\star}\ [\mathrm M_{\odot}]$ & 0.58 & 0.58 & 0.58 & 1 & fixed\\
$T_{\mathrm {eff}}$ [K] & 3792 & 3792 & 3792 & 1 & fixed\\
$L_{\star}\ [\mathrm L_{\odot}]$ & 0.7 & 0.7 & 0.7 & 1 & fixed\\
\hline\noalign{\smallskip}
\multicolumn{6}{c}{Disk parameters}\\ 
\noalign{\smallskip}\hline
$R_{\mathrm {in}}$ [AU] & 0.1 & 0.1 & 0.1 & 1 & fixed\\
$R_{\mathrm {out}}$ [AU] & 80 & 80 & 80 & 1 & fixed\\
$R_{\mathrm {gap}}$ [AU] &33 &33 &33 &1 & fixed\\
$\sigma$ [AU] &27&27&27&1&fixed\\
$\eta$ & 0.99 & 0.99 & 0.99 & 1 & fixed \\
$H_{100}$ [AU] & 7 & 5 & 10 & 6 &free\\
$\beta$ & 1.05 & 1.00 & 1.15 & 4 &free\\
$\gamma$ & 0.5 & 0.3 & 1.5 & 7 &free\\
$M_{\mathrm {dust}}$ [$\mathrm M_{\odot}$] &$10^{-3}$ & $10^{-7}$ & $10^{-2}$ & 11 & free\\
 $a^{\mathrm {mid}}_{\mathrm {max}}$ [$\mu$m] & $10^4$ & $10^3$ & $10^5$ & 5  &free\\
\hline\noalign{\smallskip}
\multicolumn{6}{c}{Observational parameters}\\ 
\noalign{\smallskip}\hline
$i$ [$\circ$] & 67.58 & 67.58 & 67.58 & 1 & fixed\\
Position angle [$\circ$] & 159.62 & 159.62 & 159.62 & 1 & fixed\\
D [pc] & 158 & 158 & 158 & 1 & fixed\\
\noalign{\smallskip}\hline
\end{tabular*}
\end{table}

\subsection{Disk model}
A flared model, which can explain the structure of protoplanetary disks \citep{2011ApJ...732...42A}, is used in this work. The disk is divided into two layers, a surface layer and an interior layer, to describe the effects of dust settling and growth. The surface layer contains small dust particles that have not grown, and large particles are present in the interior layer near the disk midplane. The density profile is
\begin{equation}
\label{rho}
\rho(R, z)\,{=}\,\frac{\Sigma(R)}{\sqrt{2 \pi} H} \exp \left(-\frac{z^{2}}{2 H^{2}}\right),
\end{equation}
where $R$ is the radial distance from the central star measured in
the disk midplane, $\Sigma(R)$ is the dust surface density, and the ${H(R)}$ is the scale height. When fitting the SED (see Section~\ref{sect:SED}), we first assume that the surface density $\Sigma(R)$ satisfies a simple power law with Gaussian depletions
\begin{equation}
\label{Sigma}
\Sigma(R)\,{=}\,\Sigma_{0} R^{-\gamma}\left\{1-\eta \exp \left[\frac{\left(R-R_{\text {gap }}\right)^{2}}{2 \sigma^{2}}\right]\right\},
\end{equation}
where $\gamma$ is the surface density gradient. The gap location and width are denoted as $R_{\mathrm {gap}}$ and $\sigma$, and are set to 32\,AU and 27\,AU according to \citet{2018ApJ...869...17L}. The parameter $\eta$ represents the depletion factor of the surface density in the gap. The disk extends from an inner to an outer radius of 0.1 and 80\,AU, respectively. The inner boundary is set to the dust sublimation radius. The outer radius is set to the location where the millimeter surface brightness within this location accounts for 90\% of the total surface brightness. The scale heights of the surface layer and the midplane are described as
\begin{equation}
\label{h}
\begin{aligned}
&H_{\mathrm{sur}}\,{=}\,H_{100}\left(\frac{R}{100 \mathrm{AU}}\right)^{\beta} \\
& {H_{\mathrm {mid}}\,{=}\,H_{\mathrm{sur}} \times H_{\mathrm{ratio}}}
\end{aligned}
\end{equation}
Considering the settling model, we assume that $85\%$ of the total dust mass is concentrated in the midplane and $15\%$ in the surface layer. The slope $\beta$ describes the extent of disk flaring. $H_{100}$ represents the scale height of the small-sized dust at 100\,AU. The scale height of the interior layer is given by $H_{\mathrm {ratio}}$, and we fixed it to 0.2, which is a typical value found by previous modeling works of protoplanetary disks \citep[e.g.,][]{2011ApJ...732...42A,Liuy2022}.

\begin{figure}
    \centering
    \includegraphics[width=0.9\columnwidth,angle=0]{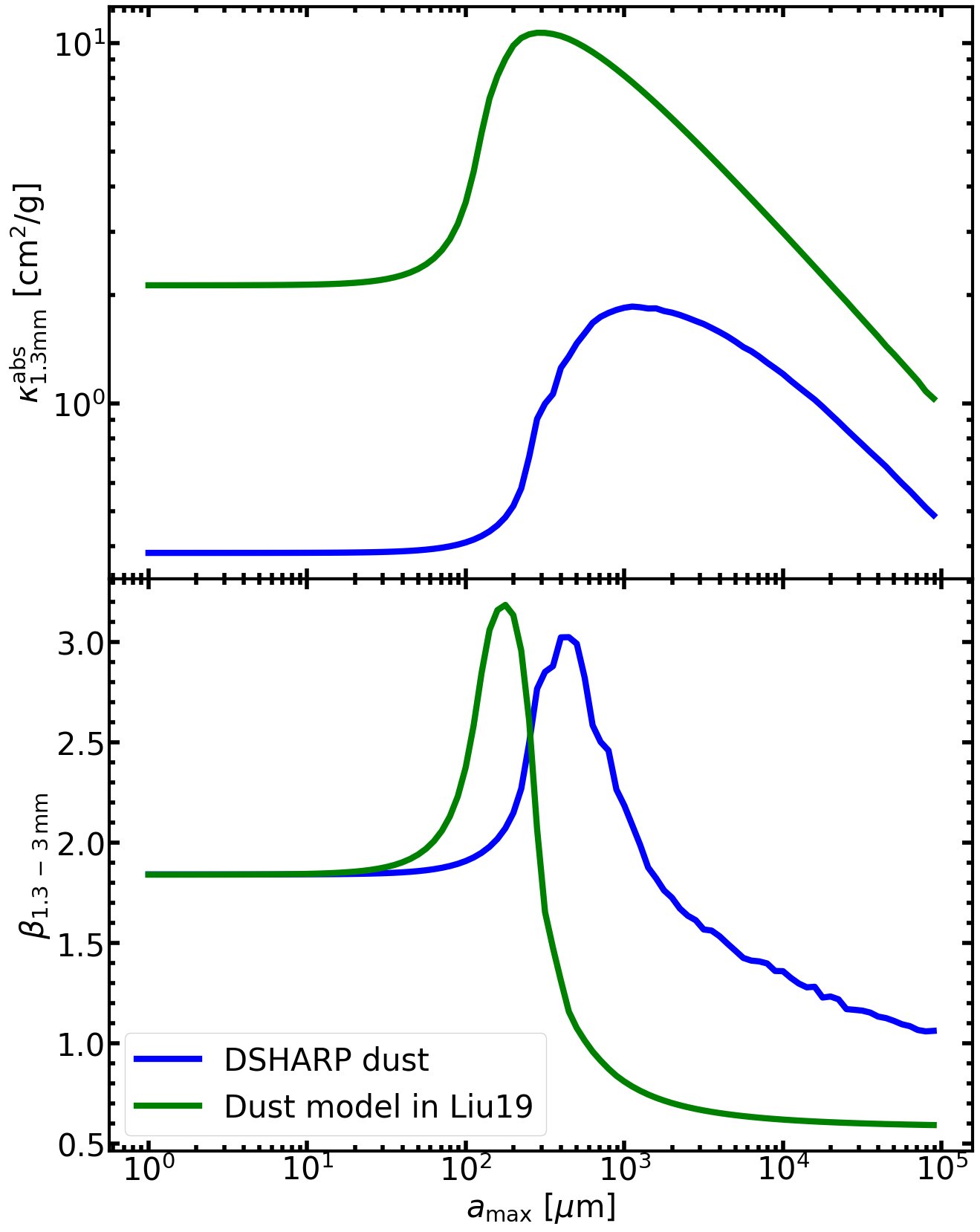}
\caption{Comparison of dust properties between different compositions. {\it Top panel:} Dust absorption coefficient at $\lambda\,{=}\,1.3\,{\rm mm}$. {\it Bottom panel:} Millimeter slope of the dust opacity within the wavelength range of [1.3\,mm, 3\,mm]. The DSHARP dust properties are indicated with blue lines, and the green lines show the dust model (75\% amorphous silicate + 25\% carbon) adopted by \citet{Liuy2019}.}
\label{fig:kappa}
\end{figure}

\subsection{Dust model}
\label{section: dust}
We assume that the dust grains are homogeneous spheres, composed of 20\% water ice, 32\% astronomical silicates, 8\% troilite, and 40\% refractory organics, with an average density of $\rho_{\mathrm {dust}}\,{=}\,1.675 \,\mathrm g/\mathrm {cm}^{-3}$. This dust prescription is introduced to interpret the ALMA data from the Disk Substructures at High Angular Resolution Project \citep[DSHARP,][]{Andrews2018}. The complex refractive indices are given by \citet{2018ApJ...869L..45B}. The grain size distribution follows the power-law ${N(a)}\,{\propto}\,a^{-3.5}, a\,{\in}\,[a_{\mathrm {min}},\ a_{\mathrm {max}}]$, where $a$ is the grain radius and ${N(a)}$ is the number of dust particles whose size is in the interval $[{a,\ a\,{+}\,da}]$. We fix the minimum dust size to $a_{\mathrm {min}}\,{=}\,0.01\,\mu $m. The maximum grain size $a_{\rm max}$ in the disk surface layer is fixed to be $1\,\mu \mathrm m$. In the midplane, we vary $a_{\mathrm {max}}$ as a function of $R$. The dielectric function and dust absorption/scattering coefficients are derived by the Bruggeman effective medium theory \citep{1935AnP...416..636B} and the MIE theory, respectively. The blue lines in Figure~\ref{fig:kappa} show the absorption coefficient at $\lambda\,{=}\,1.3\,{\rm mm}$ and the opacity slope derived within the wavelength range of $[1.3\,\rm{mm},\,3\,\rm{mm}]$ as a function of maximum grain size $a_{\rm max}$.

\subsection{Stellar heating}
We assume that the DS\,Tau disk is passively heated by absorbing energy from the central star \citep{1997ApJ...490..368C}. \citet{2018ApJ...868...28F} calculated the accretion luminosity to be $0.05\,\mathrm\,L_{\odot}$, which is only 7\% of the blackbody luminosity of the central star. Furthermore, accretion disk theory shows that only about\,{<}\,50\% of this accretion energy contributes to the heating of the inner region ($R$\,{<}\,2\,AU) of protoplanetary disks \citep{1998ApJ...509..802C}. Therefore, we ignore the effect of accretion in the simulation. \par

The central star has a spectral type of M0.4 and an effective temperature of 3800\,K \citep{2014ApJ...786...97H}. The incident stellar spectrum is taken from the Kurucz atmosphere models with solar metallicity and $\log g\,{=}\,3.5$ \citep{1994KurCD..19.....K}. The radiative transfer problem is solved self-consistently considering 160 wavelengths logarithmically distributed in the range of [0.1, 10000\,$\mu$m].
\section{Fitting approach}
\label{sect:Fit}
Fifteen parameters are used to define a particular model (see Table~\ref{parameter}). In order to reduce model degeneracies, we fixed stellar and observational  parameters, leaving out five free parameters: scale height ($H_{100}$), flaring index ($\beta$), surface density gradient index ($\gamma$), total dust mass ($M_{\mathrm {dust}}$) and maximum grain size in the disk midplane ($a^{\mathrm {mid}}_{\mathrm {max}}$). The flaring index ($\beta$) and scale height ($H_{100}$) profoundly affect the SED profile, while the total dust mass and grain size are constrained by the surface brightness at Band 6 and 3. \par 

%For fitting multi-band AlMA images, we iterate the surface density profile to accommodate the complex disk structure of DS\,Tau (see Section~\ref{sect:grid search}). The final surface density obtained from the iterative process will not affect the fitting results of SED.\par

\subsection{Fitting the SED}
\label{sect:SED}
First, we fit the SED to find a parameter set that is representative of the disk geometry. For the flaring index and scale height, we searched the ranges, $1.00\,{\leq} \beta \,{\leq}\,1.15$ and $5\,\mathrm {AU}\,{\leq}\,H_{100}\,{\leq}\,10\,\mathrm {AU}$, respectively. The parameter $\gamma$ is explored in the range from 0.3 to 1.5. For simplicity, we assume that in the SED fitting the maximum grain size in the disk midplane ($a^{\mathrm {mid}}_{\mathrm {max}}$) does not vary with $R$, and consider a range of [$10^3$, $10^5$\,$\mu \mathrm m$]. Later in Sect.~\ref{sect:grid search}, we will show that this assumption is not sufficient to reproduce the radial variation in the measured spectral index.\par

Under the optically thin condition, the total dust mass can be estimated by
\begin{equation}
\label{m}
M_{\mathrm{dust}}\,{=}\,\frac{D^{2} F_{\nu}}{\kappa_{\nu} B_{\nu}\left(T_{\mathrm{dust}}\right)},
\end{equation}
where $D$ is the distance to the object and $F_{\nu}$ is the flux density at 1.3\,mm. Using the DSHARP dust opacities described in Sect.~\ref{section: dust}, we calculate the absorption coefficient $\kappa_{\nu}\,{=}\,1.8\,\rm{cm^{2}/g}$ by assuming a maximum grain size of $1000\,\mu \rm{m}$. Adopting a dust temperature of $T_{\rm dust}\,{=}\,20\,\rm{K}$, $M_{\mathrm {dust}}$ is estimated to be $6.3\,{\times}\, 10^{-5}\,M_{\odot}$. Hence, in our parameter search, we consider a broad range of $[10^{-7}, 10^{-2}\,\mathrm M_{\odot}]$ in the logarithmic space.\par

Figure~\ref{sed} shows a comparison between the best-fit model and observed SEDs. The best-fit model is identified as the one with the minimum $\chi^2_{\rm SED}$. In order to search for a good starting point as far as possible for the subsequent fitting to the ALMA images which is the key of this study, calculating $\chi^2_{\rm SED}$ only takes the data points at the two ALMA wavelengths into account. The best-fit flaring index $\beta\,{=}\,1.05$ and scale height $H_{100}\,{=}\,7\,\rm{AU}$ are recorded in Table~\ref{parameter}. Both parameters will be fixed during the fitting to the multi-band ALMA images. Other parameters, such as the surface density gradient $\gamma$, total dust mass $M_{\mathrm {dust}}$, and grain size $a^{\mathrm {mid}}_{\mathrm {max}}$, will be further investigated.  We also checked the model that is returned from fitting the entire part of the SED, and found that all of the parameter values, except for $H_{100}=10\,{\rm AU}$, are identical to those of the best-fit model. Such a mild difference in $H_{100}$ is not expected to have a significant impact on the result of image fitting.

\subsection{Fitting the multi-band ALMA images}
\label{sect:grid search}
After obtaining representative parameters for the disk geometry, the next step is to fit the millimeter surface brightness and spectral index. Fitting the ALMA observations is performed in the image plane, rather than in the interferometric (u, v) space, because the construction of the dust surface density is directly linked to the brightness profile, see Sect.~\ref{sec:iteration}. Interferometric observations are affected by the sampling of the uv space. We checked the data, and found that the difference in the uv sampling between the two bands is not significant. The minimum and maximum baseline lengths are similar at both bands, i.e., [21\,m, 3.6\,km] and [90\,m, 8.5\,km] for band 6 and band 3, respectively. The maximum recoverable scales for band 6 and band 3 observations are ${\sim}\,1.8^{\prime\prime}$ and ${\sim}2.6^{\prime\prime}$, respectively, sufficiently larger than the disk outer radius ($R_{\rm out}\,{=}\,80\,\rm{AU}$, or ${\sim}\,0.5^{\prime\prime}$). These observational characteristics indicate that our strategy of image fitting is not likely to lead to spurious results. We use \texttt{RADMC-3D} to generate synthetic images at the two wavelengths, which are convolved with a 2D Gaussian beam with a size of [$0.13^{\prime\prime}, 0.09^{\prime\prime}$]. Then, the radial profiles of the surface brightness are obtained for the comparison with the observation.\par

Under the assumption of a constant $a^{\mathrm {mid}}_{\mathrm {max}}$ along the radial direction, the SED can be well fitted with the surface density function given in Eq.~\ref{Sigma}. However, within the whole  parameter space composed of $\gamma$, $M_{\mathrm {dust}}$ and $a^{\mathrm {mid}}_{\mathrm {max}}$, it is challenging to simultaneously fit the surface brightnesses at both ALMA bands. We think that there are two reasons for this problem. The first reason is that the surface density formula expressed as Eq.~\ref{Sigma} is not sufficient to capture the complex substructures. Secondly, it is clearly shown in panel (b) of Figure~\ref{fig:bestfit} that the spectral index $\alpha_{\rm mm}$ varies from ${\sim}\,2.1$ in the inner disk to ${\sim}\,3$ in the outer regions, which implies that $a^{\mathrm {mid}}_{\mathrm {max}}$ is not constant along the radial direction.\par

\subsubsection{Iteration for the dust surface density}
\label{sec:iteration}
We use an iterative method to construct the surface density profile for DS\,Tau. We take Eq.~\ref{Sigma}, and set $\gamma\,{=}\,0.5$ as the starting profile. 
The iteration process goes as follows:
\begin{itemize}
\item[(a)] Fixing the flaring index $\beta\,{=}\,1.05$ and scale height $H_{100}\,{=}\,7\,\rm{AU}$, we simulate the continuum image at the wavelength of 2.9\,mm. Choosing 2.9\,mm instead of 1.3\,mm is due to the fact that the disk is optically thinner at longer wavelengths. Hence, the 2.9\,mm ALMA image better traces the dust density distribution. \\
\item[(b)] Convolve the model image with the ALMA beam. \\
\item[(c)] The radial brightness profile is extracted from the convolved model image in the same way as what we have done on the ALMA image, see Sect.~\ref{sect:obs}.\\ 
\item[(d)] We divide the observed surface brightness by the model brightness at each $R$ to obtain the scaling coefficient $\xi$(R):
\begin{equation}
 \xi(R)\,{=}\,\frac{I_{\nu,\rm{obs}}(R)}{I_{\nu,\rm{mod}}(R)}
\end{equation}
\item[(e)] Multiply the surface density by the scaling coefficient $\xi$(R) to obtain a new surface density. Loopback to step (a) and the iteration continues. \par
\end{itemize}

According to our experiments, convergence can be achieved after about 12 iterations when the surface density of two adjacent iterations changes by less than $5\%$ at any radius.

\subsubsection{Grid search for the maximum grain sizes in the midplane}
In order to build the radial profile of the maximum grain size, we divide the midplane into 32 equally spaced parts with a width of 2.5\,AU in radius. Each of these 32 parts has different maximum grain sizes ($a^{\mathrm {mid}}_{\mathrm {max}}$). The width of each part is only one-eighth of the size of the ALMA beam, meaning that our setup is sufficient for resolving the disk substructures. Using more radial bins will take more computational resources, and the effect on the results is not significant. \par

Literature studies find that $a^{\mathrm {mid}}_{\mathrm {max}}$ approximately follows a power law \citep[e.g.,][]{Trotta2013,2016A&A...588A..53T,Carlos2019},
\begin{equation}
a_{\mathrm{max}}^{\operatorname{mid}}(R)=a_{\max 0}\left(\frac{R}{R_{0}}\right)^{-b}
\label{eq:amax}
\end{equation}
where $R_{0}\,{=}\,40\,\rm{AU}$ is the characteristic radius, and $a_{{\max 0}}$ is a reference grain size. We first take such a distribution, and conduct an extensive search for the two free parameters $a_{\max0}$ and $b$ within the ranges of [100, $10000\,\mu{\rm m}$] and [1, 8], respectively. The best model ($a_{\max0}\,{=}\,1000\,\mu{\rm m}$, $b\,{=}\,6$) cannot simultaneously fit the two ALMA images, implying that the maximum grain size in the midplane of DS\,Tau's disk cannot be represented by a simple power-law function. Starting from the best model returned from such a parameter study, we further use the grid search method to adjust $a^{\mathrm {mid}}_{\mathrm {max}}$ in each of the 32 parts to improve the fitting to the observed spectral indices $\alpha_{\mathrm {mm}}(R)$.  \par

The grid search for $a^{\mathrm {mid}}_{\mathrm {max}}$ is performed in each of the 32 radial bins, from the inner region gradually to the outer disk. The initial value for $a^{\mathrm {mid}}_{\mathrm {max}}$ in each bin is set according to Eq.~\ref{eq:amax} with $a_{\max0}\,{=}\,1000\,\mu{\rm m}$, $b\,{=}\,6$. Taking a particular radial bin as an example, we create a dense grid of models with $a^{\mathrm {mid}}_{\mathrm {max}}$ logarithmically sampled around the initial value in that bin, while $a^{\mathrm {mid}}_{\mathrm {max}}(s)$ of other radial bins remain to their initial choices. The spectral indices of these models are calculated from the simulated images at 1.3 and 2.9\,mm, and fitted to the observation. The grain size in this radial bin is determined to be that of the best-fit model. Meanwhile, the so-called initial value for $a^{\mathrm {mid}}_{\mathrm {max}}$ in this particular bin is updated, because it is related to the search for $a^{\mathrm {mid}}_{\mathrm {max}}$ in other bins. The maximum grain size in the next radial bin is fitted in the same way. After all of the 32 bins have been explored, we obtain a radial profile of grain size in the midplane. We note that for each model established in the process of grid search, the iteration for the dust surface density $\Sigma( R)$ described in Sect.~\ref{sec:iteration} is carried out from scratch, ensuring that all the radiative transfer models are fully independent and self-consistent. This is the most significant difference between our approach and the methods used in \citet{2016A&A...588A..53T} and \citet{2021ApJS..257...14S}, see Sect.~\ref{sect:intro}. 

The $a^{\mathrm {mid}}_{\mathrm {max}}$ for $R\,{<}\,20\,\mathrm{AU}$ are fixed to $10\,\mathrm {cm}$. After many simulation tests, we found that this inner region is optically thick even at 2.9\,mm, indicating that the grain size can no longer be constrained by the observed spectral index. On the one hand, grain sizes below $10\,\mathrm {cm}$ result in spectral indices that deviate further from the observational values. On the other hand, if the grain size is larger than 10\,cm, it may exceed the applicable range of MIE theory. Moreover, beyond 10\,cm the effect of increasing $a^{\mathrm {mid}}_{\mathrm {max}}$ on the result is minor. For these reasons, we make the choice of 10\,cm for $a^{\mathrm {mid}}_{\mathrm {max}}$ for $R\,{<}\,20\,\mathrm{AU}$. 
%After the best-fit grain sizes are found, we separately vary $a^{\mathrm {mid}}_{\mathrm {max}}$ from the best-fit value in each bin. The uncertainties of $a^{\mathrm {mid}}_{\mathrm {max}}$ are determined from
%After the best-fit grain sizes are found, the uncertainties of $a^{\mathrm {mid}}_{\mathrm {max}}$ for each bin (within $20\,{<}\,R\,{<}\,80\,\mathrm {AU}$) are determined from
After the best-fit solution is obtained, for each radial bin we simulated 17 new models with the maximum grain size $a_i$ centered at the best-fit value $a_{\mathrm {best-fit}}$ and logarithmically spaced from $\mathrm {log}(a_{\mathrm {best-fit}})\,{-}\,0.4$ to  $\mathrm {log}(a_{\mathrm {best-fit}})\,{+}\,0.4$. The uncertainties are estimated from
\begin{equation}
\label{error}
\sigma^2\,{=}\,\left[\frac{\sum_{i}^{N}\omega_{i}\left(\mathrm {log} (a_{i})\,{-}\,\mathrm {log}(a_{\mathrm {best-fit}})\right)^{2}}{\sum_{i}^{N}\omega_{i}}\right]\,{\times}\,\frac{N}{N-1},
\end{equation}
where the weighting factor equals to one over the chi-square of the model $\omega_{i}\,{=}\,\frac{1}{\chi_{i}^2}$, and $N\,{=}\,17$ is the number of models. 

%The derived dispersions are used as uncertainties that are shown in Figure~\ref{alpha-r}. We separately vary maximum grain size, $a_{\mathrm i}$, from the best-fit value, $a_{\mathrm {best\ fit}}$, in each bin and the total number of models modified in one bin is 

%\begin{figure}[htbp]
%\centering
%\includegraphics[width=\columnwidth, angle=0]{toomre.png}
%\caption{The Toomre Q of the disk along the radial direction. The horizontal black dashed line show the value of %Toomre Q parameter equals 1.}
%\label{toomre}
%\end{figure}

\section{Results and discussion}
\label{sect:discussion}

\subsection{Overview of the modeling results}

The results of fitting to the surface brightnesses and spectral indices, together with the retrieved $a^{\mathrm {mid}}_{\mathrm {max}}\,{-}\,R$ profile, are presented in panels (a), (b) and (c) of Figure~\ref{fig:bestfit}, respectively. As can be seen, the best-fit model well reproduces the observation for $R\,{\ge}\,20\,\rm{AU}$. We find that the grain size increases 
from ${\sim}\,30\,\mu \rm m$ in the outer disk to centimeters in the inner regions. This trend is consistent with theories of dust evolution in protoplanetary disks: small dust grains are tightly coupled to the gas and are distributed farther away, whereas large pebbles drift inward \citep{1977MNRAS.180...57W,2008A&A...480..859B,Birnstiel2016}. The $a^{\mathrm {mid}}_{\mathrm {max}}\,{-}\,R$ profile exhibits some fluctuations on spatial scales smaller than the beam size. Future high-resolution observations are desired to confirm their reality.\par

\begin{figure}
\centering
\includegraphics[width=\columnwidth,angle=0]{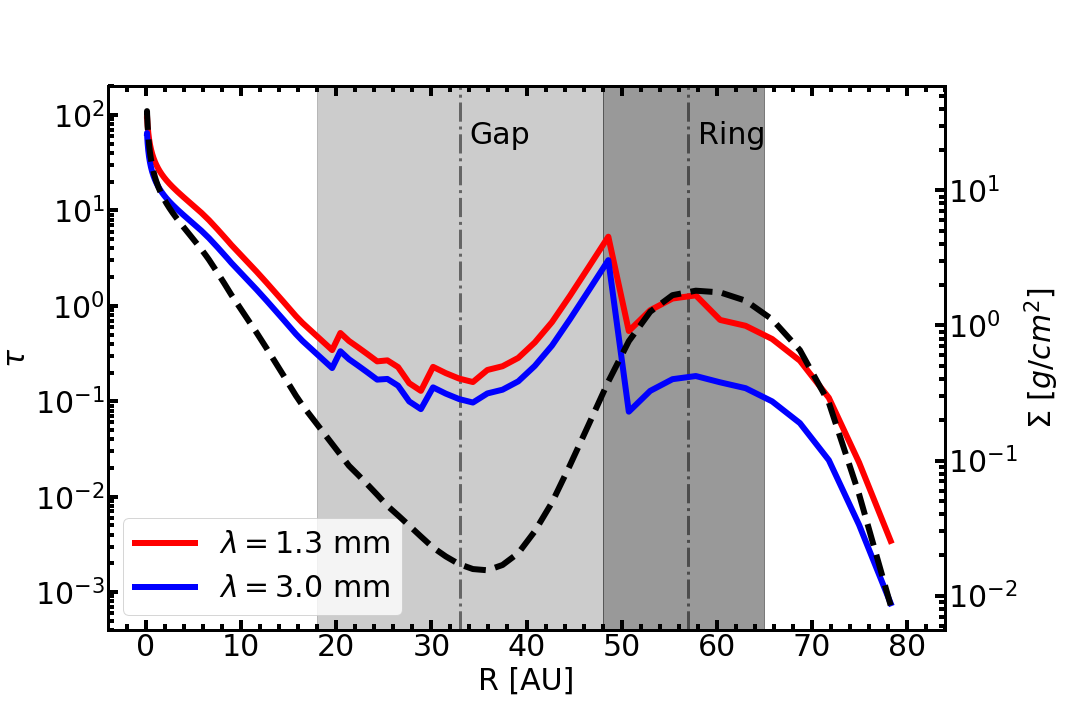}
\caption{Optical depth of the best-fit model at the wavelength of 1.3\,mm (red solid line) and 2.9\,mm (blue solid line). The surface density profile is shown as the black dashed line.}
\label{fig:tau}
\end{figure}

The constructed surface density and optical depth at 1.3 and 2.9\,mm are shown in Figure~\ref{fig:tau}. The optical depth increases from the outer disk to inner regions. The bulk of the disk, except for $R\,{<}\,20\,{\rm AU}$ and $R\,{\sim}\,50\,{\rm AU}$, is optically thin. The calculation of optical depth is related to the dust density and extinction coefficient (controlled by the grain size in our study). Although the surface density in the ring is larger than that at the boundary ($R\,{\sim}\,50\,{\rm AU}$) between the gap and ring, the grain size at the boundary is much larger than that in the ring center, and so does the dust opacity, see panel (c) of Figure~\ref{fig:bestfit} and the top panel of Figure~\ref{fig:kappa}. Consequently, the optical depth at $R\,{\sim}\,50\,{\rm AU}$ shows a spike peaking around 3. 

Theoretically, disk substructures may affect the growth of dust grains. At the outer edge of the disk, the grain size is about $30\,\mu{\rm m}$, and it does not increase significantly in the ring ($R\,{=}\,57\,{\rm AU}$). However, dust grains grow up by two orders of magnitude at the boundary between the gap and ring. We run additional models to confirm the necessity of such a jump in grain size for fitting the observation, see Figure~\ref{fig:moremodel1}. There are two possible mechanisms for a steep increase in grain size. The first scenario is due to the rapid grain growth around snow lines. As the temperature decreases along the radius, some of the volatile molecules reach their freezing point, and condense on the surface of dust grains at specific locations. The ice mantles change the effective strength of the dust particles and the results of their collisions when dust particles cross the snow line \citep{2010A&A...516L..14B,2011ApJ...728...20S,Zhang2015}. If the collision velocity is less than the fragmentation velocity of the particles, the particles may grow and drift rapidly into the inner disk. On the contrary, the particles will break up and become smaller and drift slower, similar to traffic congestion. However, according to our model, the midplane temperature at the boundary between the gap and ring is about 10\,K which is not associated with condensation fronts of main volatile species in protoplanetary disks.\par

Secondly, a pressure bump may exist at a location where the grain size changes sharply. The pressure gradient in the disk is generally decreasing but not necessarily monotonically. There may be a local pressure maximum at a particular location, called a pressure bump. Dust grains at the pressure maximum are trapped because the radial drift velocity of the grains is slowed or even stopped. The accumulation of dust particles promotes the process of grain growth, resulting in a sharp increase in grain size. The cause of the pressure bumps may be due to an embedded planet in the vicinity of the gap. Through hydrodynamic simulations, \citet{2020MNRAS.495.1913V} proposed that a planet with a mass of $M_{\mathrm P}\,{=}\,3.5\,{\pm}\,1\,\mathrm{M_{\mathrm {Jup}}}$ located at $R\,{\sim}\,35\,\rm{AU}$ can produce the observed gap and ring structures in the DS\,Tau disk. 

Numeric simulations of dust trapping due to planets have demonstrated that larger dust grains are more concentrated toward the ring center \citep[e.g.,][] {Pinilla2012,Pinilla2015a,Pinilla2015b}, which appears inconsistent with our finding that the jump in grain size occurs in the interface region between the gap and the ring. In Figure~\ref{fig:moremodel1}, we have shown that larger dust grains trapped in the ring center will produce spectral indices contradictory with the observations. The same inconsistency is also seen in Figure 7 of \citet{2020MNRAS.495.1913V}. These imply that based on the current data, the majority of larger dust grains are probably not located in the ring center of the DS\,Tau disk. However, the measured spectral indices actually trace the mean dust properties over a radius range depending on the beam size, i.e., ${\sim}\,20\,\rm{AU}$ in our case. Future higher resolution observations, especially carried out at longer wavelengths, are needed to better characterize the variation of spectral index on small scales, and therefore to place stronger constraints on the grain size and its correlation with disk substructures.

\subsection{Potential for forming a ${\sim}\,3.5\,{M_{\rm Jup}}$ mass planet}

We estimate how much material could have been depleted within the gap in order to roughly evaluate the potential for forming the planet inferred by \citet{2020MNRAS.495.1913V}.  For this purpose, we build a smooth surface density profile without substructures on the basis of the best-fit surface density. The surface densities for $6\,{\le}\,{R}\,{\le}\,60\,{\rm AU}$ are parameterized by a power law that connects the best-fit surface densities at $R\,{=}\,6\,\rm{AU}$ and $R\,{=}\,60\,\rm{AU}$, and the surface densities outside the radius range of [6\,AU, 60\,AU] are set to the best-fit values. Integrating the smooth surface density gives the total dust mass prior to the formation of the planet. The difference between this mass and the best-fit dust mass is ${\sim}\,2\,\mathrm M_{\mathrm {Jup}}$, which is the amount of material that could contribute to the planet formation. Taking the gas-to-dust mass ratio (typically ${\sim}\,30$ in protoplanetary disks) into account, the depleted mass is sufficient for forming a $3.5\,\mathrm M_{\mathrm {Jup}}$ mass planet. \par

\subsection{Gravitational instability}

We check whether the disk is gravitationally stable by using the Toomre $Q$ criterion
\begin{equation}
Q=\frac{c_{\mathrm{s}} \Omega_{\mathrm{K}}}{\pi G \Sigma_{\mathrm{g}},
\label{eq:toomre}}
\end{equation}
where $\Omega_{\mathrm{K}}\,{=}\,\sqrt{G M_{\star}/R^3}$ is the Keplerian angular velocity, and $G$ is the gravitational constant \citep{1964ApJ...139.1217T}. The sound speed $c_{\mathrm {s}}\,{=}\,\sqrt{k_{\mathrm{B}} T / \mu_{m} m_{\rm{H}}}$ is calculated from the midplane temperature $T$, with a mean molecular weight $\mu_{m}\,{=}\,2.34$. The quantities $k_{B}$ and $m_{H}$ are the Boltzmann constant and the hydrogen mass, respectively. To have the gas surface density $\Sigma_{\mathrm{g}}$, we simply scale the best-fit dust surface density by a gas-to-dust mass ratio of 30 that is assumed to be uniform in the entire disk. 

The $Q$ value is greater than unity at all radii, except for the ring region with $Q\,{\sim}\,0.5$. As shown by numeric simulations of dust trapping in rings, the gas-to-dust mass ratio varies as dust particles concentrate toward pressure maxima \citep[e.g.,][]{Gonzalez2017}, and is also dependent on the input stellar mass, turbulence level, fragmentation velocity, and other parameters \citep[e.g.,][]{Pinilla2020}. The local gas-to-dust mass ratio in rings can decrease down to 10, and even to unity. These theoretical predictions have also been supported by ALMA observations \citep[e.g.,][]{Ansdell2016,Zhang2021}. If a gas-to-dust mass ratio of ${<}\,10$ is assumed, the $Q$ value of the entire DS\,Tau disk will be above unity, meaning that gravitational instability will not operate.

\subsection{Effects of dust opacities}

The results outlined above are based on the modeling using the DSHARP dust opacities. Observational studies, mostly relied on infrared spectra so far, have revealed that the dust composition varies among disks \citep[e.g.,][]{Juhasz2010}. The absorption coefficient and millimeter opacity slope differ between different dust compositions. To set up the radiative transfer model for MWC\,480, \citet{Liuy2019} assumed that the dust ensemble is composed of 75\% amorphous silicate \citep{Dorschner1995} and 25\% carbon \citep{Jager1998}. As shown with green lines in Figure~\ref{fig:kappa}, the mass absorption coefficient of their dust model is systematically larger than that of the DSHARP dust prescription. Moreover, the location of the peaks as well as the steepness of the $\beta_{1.3-3\,\rm{mm}}\,{-}\,a_{\rm max}$ relation are different between the two dust models. 

We rerun the fitting procedure from scratch using the dust model chosen by \citet{Liuy2019}, with the results presented in panels (d)$-$(f) of Figure~\ref{fig:bestfit}. As can be seen, the overall shape of the $a_{\rm max}^{\rm mid}\,{-}\,R$ profile remains similar. The tendency of an inside-out decreasing grain size is unchanged. However, the grain sizes within the radius range from 20\,AU to 50\,AU are systematically smaller than those retrieved using the DSHARP dust model. Consequently, the originally claimed jump becomes a bump at the boundary between the gap and the ring, and the amplitude of the bump is roughly one order of magnitude. We also run an additional model to check the necessity of such a discontinuity in the radial grain size profile. The result indicates that the model cannot reproduce the spectral indices at the corresponding radial locations if the bump is flattened out, see Figure~\ref{fig:moremodel2}.

%The planet's properties affect the depth of the gap in the protoplanetary disk, so we can estimate the planet's mass by analyzing the depth of the gap. The mass of the dust in the disk obtained from the surface density %calculation is $1.4\,{\times}\, 10^{-3}\,\mathrm M_{\odot}$. The surface density at the gap location drops sharply and is one-hundredth of the surface density at the ring. Based on the calculations, we estimate that %the gap has lost about 2\,$\mathrm M_{\mathrm {Jup}}$ of material. This mass is close to the planetary mass of $M_{\mathrm P}\,{=}\,3.5\,\mathrm M_{\mathrm {Jup}}$ derived from the \citet{2020MNRAS.495.1913V} %simulation. The depleted mass is not enough to form the simulated planetary mass because the planet's mass does not fully come from the dust in the gap. The planet accretes some mass in the outer disk, then migrates %into the gap location and gradually gains more mass. However, we only calculate the mass loss of the gap, which may underestimate the planet's mass. The specific properties of the planet remain to be observed at high %resolution in the future.

\section{Summary}
\label{sect:summary}
In this work, for the first time, we build self-consistent radiative transfer models for the DS\,Tau disk to simultaneously fit its SED and millimeter surface brightnesses at two ALMA bands. The main goal is to constrain the grain size and its variation in the radial direction. Our main conclusions are given as follows.\par
\begin{itemize}
 \item [1)] The inner disk ($R\,{<}\,20\,\rm{AU}$) is optically thick even at 3\,mm. The gap is optically thin. However, the optically depth is higher than unity at the junction between the gap and the ring. 
 
  \item[2)] We find that the grain sizes range from 10\,cm in the inner disk down to ${\sim}\,30\,\mu{\rm m}$ in the outer regions. This outside-in increasing tendency of grain size is consistent with theories of dust evolution in protoplanetary disks. 
  \item [3)] When the DSHARP dust opacities are adopted, the grain size varies by two orders of magnitude at the interface region between the gap and the ring. Such a discontinuity in the radial grain size profile is also observed when a different dust model is used in the radiative transfer analysis, though its amplitude is reduced. Nevertheless, limited by the spatial resolution of current data, future observations at higher angular resolutions are needed to better constrain the grain size and its variation in the vicinity of disk substructures.
  
  \item [4)] Assuming that the gap is produced by an embedded planet, we calculated the missing dust mass within the gap. That is ${\sim}\,2\,M_{\rm Jup}$. By taking the gas-to-dust mass ratio into account, this mass is enough to form a planet with a mass of $M_{\rm p}\,{=}\,3.5\,M_{\rm Jup}$ inferred from literature hydrodynamic simulations of planet-disk interaction for DS\,Tau. Future high resolution and high contrast observations are required to confirm the existence/absence of such a planet.
\end{itemize}

\section*{Acknowledgements}
We thank the anonymous referee for constructive comments that highly improved the manuscript. YL acknowledges the financial support by the National Natural Science Foundation of China (Grant No. 11973090), and the science research grants from the China Manned Space Project with NO. CMS-CSST-2021-B06. HW acknowledges the financial support by the National Natural Science Foundation of China (Grant No. 11973091). YW acknowledges the support by the National Natural Science Foundation of China (Grant No. 11873094 and 12041305) and the support by the Natural Science Foundation of Jiangsu Province (Grants No. BK20221163). We thank Feng Long, Min Fang and Benedetta Veronesi for insightful discussions. {\it PACS} has been developed by a consortium of institutes led by MPE (Germany) and including UVIE (Austria); KU Leuven, CSL, IMEC (Belgium); CEA, LAM (France); MPIA (Germany); INAF-IFSI/OAA/OAP/OAT, LENS, SISSA (Italy); IAC (Spain). This development has been supported by the funding agencies BMVIT (Austria), ESA-PRODEX (Belgium), CEA/CNES (France), DLR (Germany), ASI/INAF (Italy), and CICYT/MCYT (Spain). ALMA is a partnership of ESO (representing its member states), NSF (USA) and NINS (Japan), together with NRC (Canada), MOST and ASIAA (Taiwan), and KASI (Republic of Korea), in cooperation with the Republic of Chile. The Joint ALMA Observatory is operated by ESO, AUI/NRAO and NAOJ. This work has made use of data from the European Space Agency (ESA) mission Gaia (https://www.cosmos.esa.int/gaia), processed by the Gaia Data Processing and Analysis Consortium (DPAC, https://www.cosmos.esa.int/web/gaia/dpac/consortium). Funding for the DPAC has been provided by national institutions, in particular the institutions participating in the Gaia Multilateral Agreement.

%%%%%%%%%%%%%%%%%%%%%%%%%%%%%%%%%%%%%%%%%%%%%%%%%%
\section*{Data Availability}
% The flux densities of DS\,Tau used in this work are available in \cite{2018yCat.1345....0G}, \cite{1995ApJS..101..117K}, \cite{2003yCat.2246....0C}, \cite{2012yCat.2311....0C}, \cite{2010ApJS..186..111L}, \cite{2017ApJ...849...63R}, \cite{2005ApJ...631.1134A}, \cite{1995ApJ...439..288O}, \cite{2018ApJ...869...17L}, \cite{2020ApJ...898...36L}, \cite{2010A&A...512A..15R}. The ALMA observation and data reduction of DS\,Tau can be found in \citet{2018ApJ...869...17L,2020ApJ...898...36L}.
% The RADMC-3D code package is publicly available at https://github.com/dullemond/radmc3d-2.0.
The data underlying this article will be shared on reasonable request to the corresponding author.



\bibliographystyle{mnras}
\bibliography{bibtex.bib} % if your bibtex file is called example.bib

%\begin{thebibliography}{99}
%\end{thebibliography}

%%%%%%%%%%%%%%%%%%%%%%%%%%%%%%%%%%%%%%%%%%%%%%%%%%

%%%%%%%%%%%%%%%%% APPENDICES %%%%%%%%%%%%%%%%%%%%%

\appendix

\section{More radiative transfer models}
\begin{figure}
\centering
\includegraphics[width=0.9\columnwidth,angle=0]{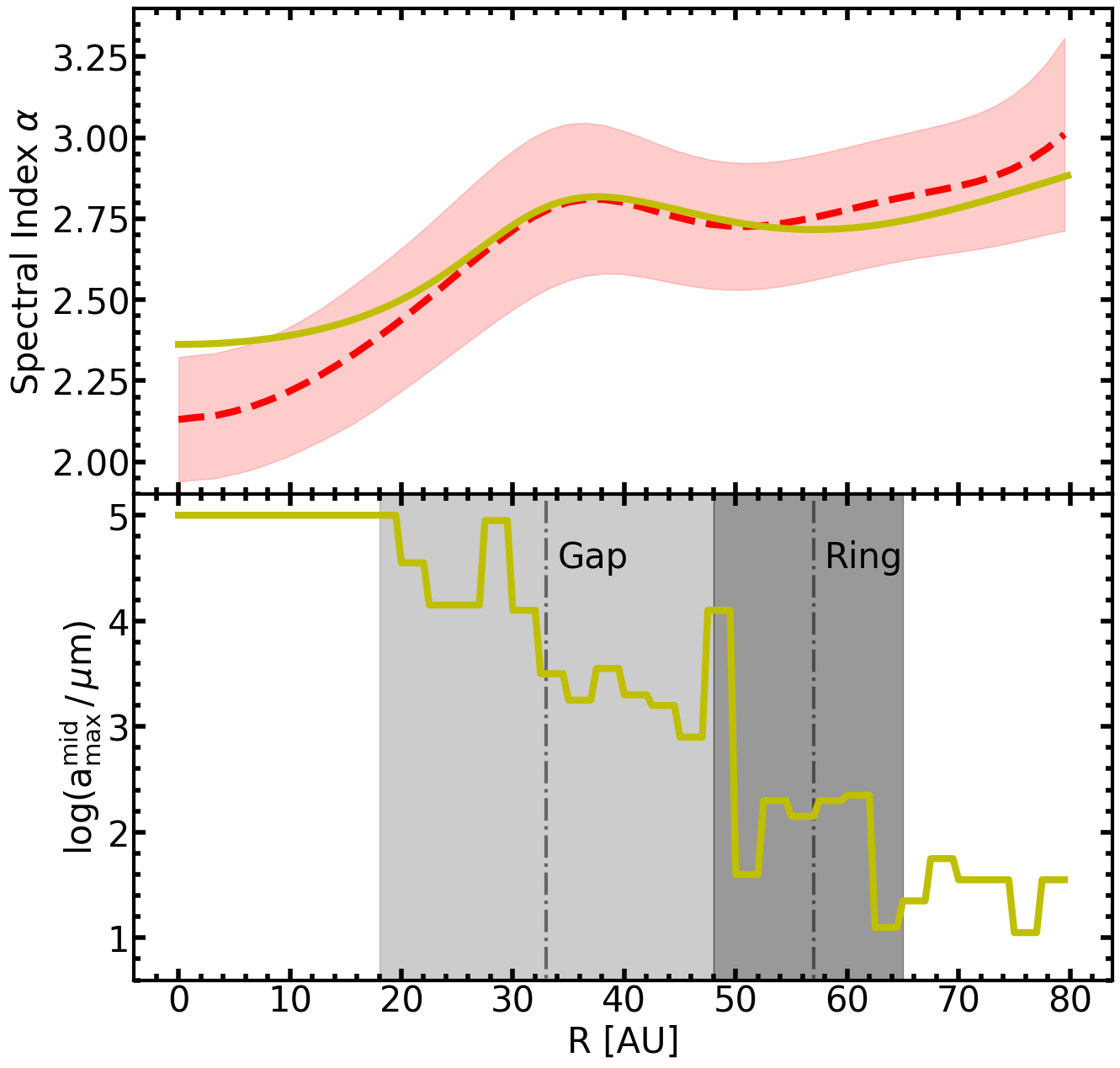}
\caption{Same as in panels (b) and (c) of Figure~\ref{fig:bestfit}, but a $10\%$ flux calibration uncertainty at both ALMA bands is taken into account in the fitting process. The DSHARP dust opacities are used to set up the radiative transfer model.}
\label{fig:calibration}
\end{figure}

In the fitting procedure, we did not consider the absolute flux calibration accuracy. We rerun the modeling by assuming a ${\sim}\,10\%$ flux calibration accuracy at both ALMA bands. The results are presented in Figure~\ref{fig:calibration}. As can be seen, the retrieved profile for the grain size is similar to that shown in panel (c) of Figure~\ref{fig:bestfit}, meaning that the absolute flux calibration uncertainty does not have a significant impact to our study. 

\begin{figure}
\centering
\includegraphics[width=0.9\columnwidth,angle=0]{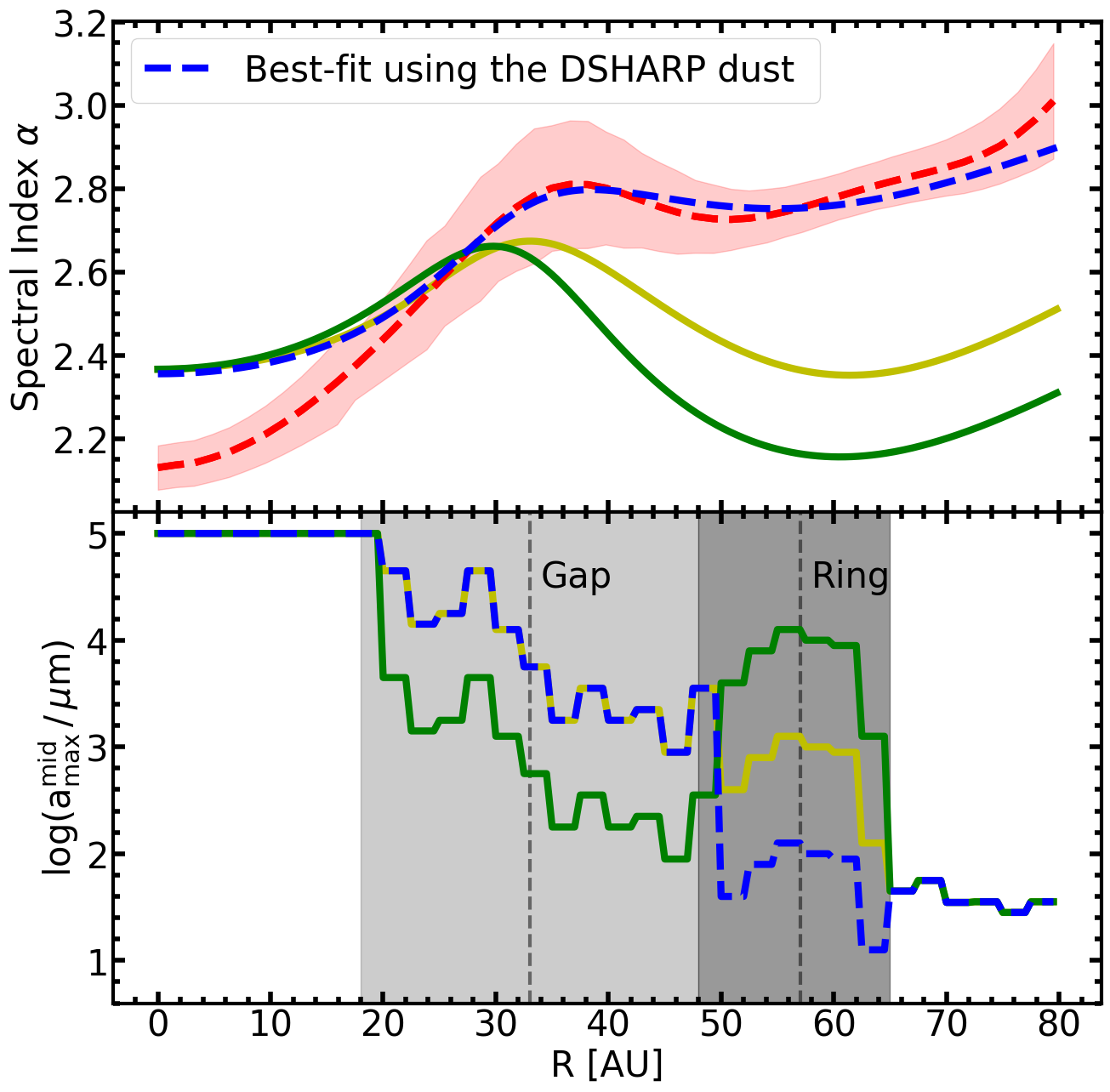}
\caption{Same as in panels (b) and (c) of Figure~\ref{fig:bestfit}, but for two different radial profiles of  $a_{\rm max}^{\rm mid}$. The yellow lines represent a relatively smooth profile of $a_{\rm max}^{\rm mid}$ (bottom panel) and the predicted spectral indices (top panel). The green lines show the result for a model in which $a_{\rm max}^{\rm mid}(s)$ in the ring are larger than those in the gap. For a better presentation, the best-fit model is shown with the blue dashed curves. Note that the DSHARP dust opacities are used to set up these models.}
\label{fig:moremodel1}
\end{figure}

As described in Sect.~\ref{sect:discussion} and shown in panel (c) of Figure~\ref{fig:bestfit}, the best-fit model has a significant jump in $a_{\rm max}^{\rm mid}$ at the boundary between the gap and ring. Moreover, $a_{\rm max}^{\rm mid}(s)$ in the gap are larger than those in the ring.  We run two additional models to check how the fitting result would be when the radial profile of $a_{\rm max}^{\rm mid}$ is different from the best fit. One model features a relatively smooth profile of $a_{\rm max}^{\rm mid}$, i.e., there is no jump in $a_{\rm max}^{\rm mid}$ in the border between the gap and ring. The resulting spectral indices are shown with the yellow solid line in the top panel of Figure\,\ref{fig:moremodel1}. In the other model, $a_{\rm max}^{\rm mid}(s)$ in the ring are devised to be larger than those in the gap. The result is indicated with the green solid line. It is clear that both setups cannot reproduce the observation.

\begin{figure}
\centering
\includegraphics[width=0.9\columnwidth,angle=0]{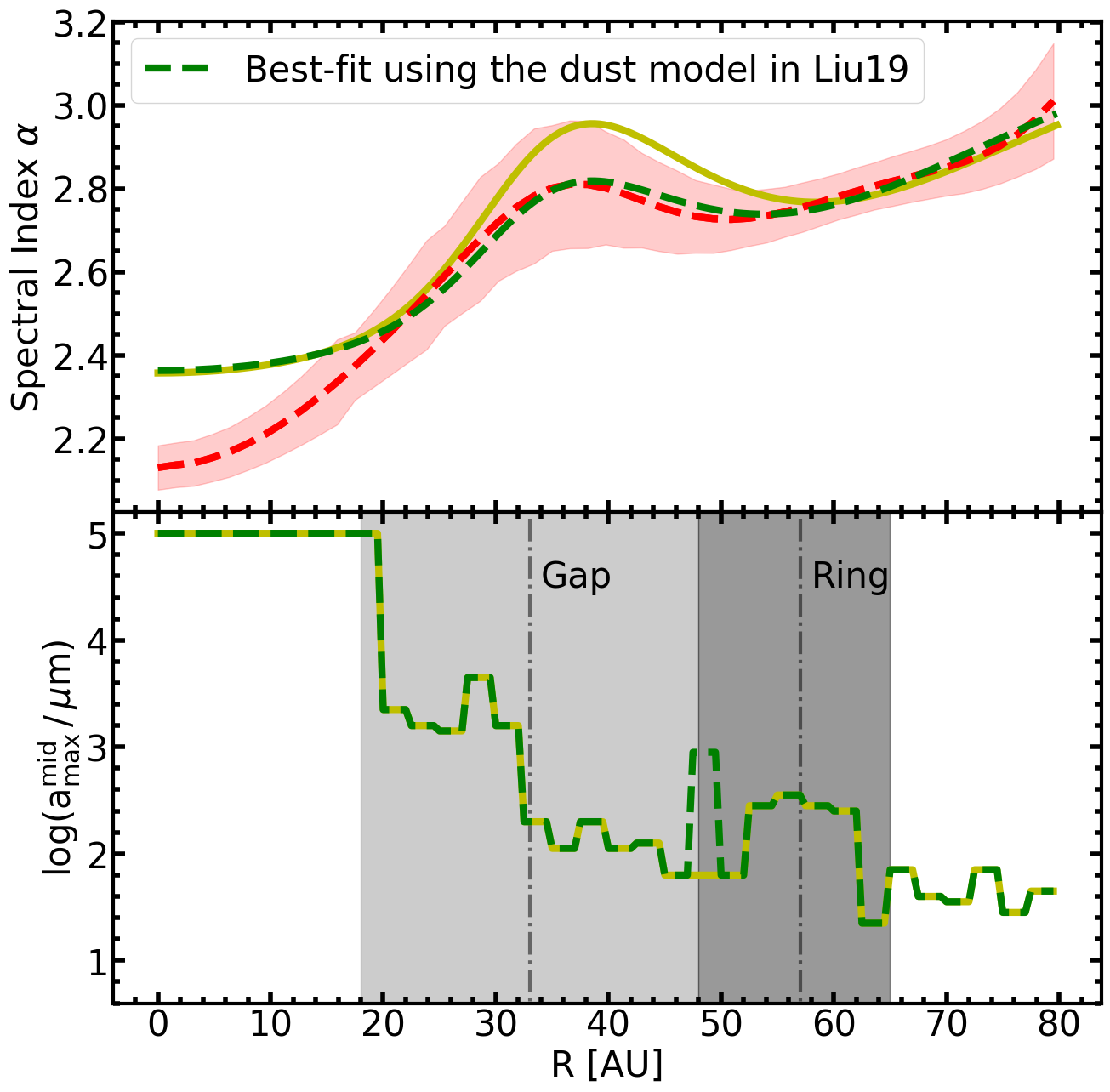}
\caption{Same as in panels (e) and (f) of Figure~\ref{fig:bestfit}, but for a radial profile of $a_{\rm max}^{\rm mid}$ in which the bump is smoothed out. For a better presentation, the best-fit model is shown with the green dashed curves. The yellow lines represent a relatively smooth profile of $a_{\rm max}^{\rm mid}$ (bottom panel) and the predicted spectral indices (top panel). Note that the dust model used in \citet{Liuy2019} is taken to run the simulation.}
\label{fig:moremodel2}
\end{figure}

As shown in panel (f) of Figure~\ref{fig:bestfit}, the best-fit model using the dust composition adopted by \citet{Liuy2019} has a bump in $a_{\rm max}^{\rm mid}$ at the boundary between the gap and ring. We run an additional model in which the bump in the $a_{\rm max}^{\rm mid}$ profile is smoothed out. Figure~\ref{fig:moremodel2} show that the new model cannot well reproduce the observed spectral indices.

%%%%%%%%%%%%%%%%%%%%%%%%%%%%%%%%%%%%%%%%%%%%%%%%%%

% Don't change these lines
%\bsp	% typesetting comment
\label{lastpage}
\end{document}